\documentclass[10pt,a4paper]{article}

\usepackage{a4}
\usepackage{latexsym, array, enumerate}
\usepackage{amstext, amsmath, amsbsy, amsthm}
\usepackage{amsfonts}
\usepackage{verbatim}
\usepackage{setspace}

\usepackage{amsmath,epsfig}
\def\R{\mathbb{R}}
\def\th{\mbox{\boldmath $\theta$}}

\def\Sig{\mbox{\boldmath $\Sigma$}}

\def\eps{\mbox{\boldmath $\epsilon$}}
\def\mmu{\mbox{\boldmath $\mu$}}
\def\s{{\bf s}}

\title{Bayesian Computation and Model Selection in Population Genetics}
\author{
  \: Christoph Leuenberger,\footnote{These two authors contributed equally to this work} \footnote{Ecole d'ing\'enieurs de Fribourg, Bd. de P\'erolles 80, 1705 Fribourg, Switzerland, christoph.leuenberger@eif.ch}
  \: Daniel Wegmann,\footnotemark[1] \footnote{University of Berne, Computational and Molecular Population Genetics Laboratory,
  3012 Berne, Switzerland, daniel.wegmann@zoo.unibe.ch}
  \: Laurent Excoffier\footnotemark[3]
}

\date{}

\begin{document}

\maketitle

\begin{abstract}
Until recently, the use of Bayesian inference in population genetics was limited to a few cases because for many realistic population genetic models the likelihood function cannot be calculated analytically .
The situation changed with the advent of likelihood-free inference algorithms, often subsumed under the term Approximate Bayesian Computation (ABC). A key innovation was the use of a post-sampling regression adjustment, allowing larger tolerance values and as such shifting computation time to realistic orders of magnitude \cite{Be}. Here we propose a reformulation of the regression adjustment in terms of a General Linear Model (GLM). This allows the integration into the sound theoretical framework of Bayesian statistics and the use of its methods, including model selection via Bayes factors. We then apply the proposed methodology to the question of population subdivision among western chimpanzees {\it Pan troglodytes verus}.

\end{abstract}

\subsection*{Introduction}

With the advent of ever more powerful computers and the refinement
of algorithms like MCMC or Gibbs sampling, Bayesian statistics has
become an important tool for scientific inference during the past
two decades. Until recently many scientists shunned Bayesian methods
-- mainly because of the philosophical problems related to the
choice of prior distributions -- but the development of hierarchical
and empirical Bayes turned them into an alternative even for
hard-core frequentists (see e.g. \cite{Ro} for a discussion of these
issues).

Consider a model ${\cal M}$ creating data ${\cal D}$ (DNA sequence
data, for example) determined by parameters $\th$ from some (bounded) parameter space $\Pi \subset \R^m$
whose joint prior density we denote by $\pi({\bf \th})$. The quantity of interest is
the posterior distribution of the parameters which can be calculated
by Reverend Bayes' golden rule

$$
\pi(\th |{\cal D})=c \cdot f_{\cal M}({\cal D}|\th) \pi(\th),
$$

where $f({\cal D}|\th)$  is the likelihood of the data and $c$ is a
normalizing constant. Direct use of this formula, however, is often
thwarted by the fact that the likelihood function cannot be
calculated analytically for many realistic population genetic models. In
these cases one is obliged to have recourse to stochastic
simulation. Tavar\'e {\it et al.} \cite{Ta} propose a rejection
sampling method for simulating a posterior random sample where the full
data ${\cal D}$ is replaced by a summary statistics ${\it s}$ (like
the number of segregating sites in their setting). Even if the
statistics are not sufficient for ${\cal D}$ -- that is, the
statistics do not capture the full information contained in the data
--, rejection sampling allows for the simulation of approximate
posterior distributions of the parameters in question (the scaled
mutation rate in their model). This approach was extended to
multiple-parameter models with multivariate summary statistics
$\s=(s_1,\dots,s_n)^T$ by Weiss and von Haeseler \cite{Wei}. In their
setting a candidate vector $\th$ of parameters is simulated from a
prior distribution and is accepted if its corresponding vector of
summary statistics is sufficiently close to the observed summary
statistics $\s_{obs}$ with respect to some metric in the space of
$\s$, i.e. if $\mbox{dist}(\s,\s_{obs})<\epsilon$ for a fixed
tolerance $\epsilon$. If we suppose that the likelihood $f_{\cal M}(\s|\th)$ of the full model is continuous
and non-zero around $\s_{obs}$ then the likelihood of this truncated model ${\cal M}_\epsilon(\s_{obs})$ obtained by this accept-reject process is given by
\begin{equation}\label{truncated likelihood}
    f_\epsilon(\s|\th)= \mbox{Ind}(\s \in {\cal B}_\epsilon(\s_{obs}))
    \cdot f_{\cal M}(\s|\th)\cdot(\int_{{\cal B}_\epsilon} f_{\cal M}(\s|\th)d\s)^{-1}
\end{equation}
where ${\cal B}_\epsilon={\cal B}_\epsilon(\s_{obs})=\{\s \in \R^n|\mbox{dist}(\s,\s_{obs})<\epsilon\}$ is the
$\epsilon$-ball in the space of summary statistics and $\mbox{Ind}(\cdot)$ is the indicator function.
Observe that $f_\epsilon(\s|\th)$ degenerates to a (Dirac) point measure centered at $\s_{obs}$ as
$\epsilon \rightarrow 0$. If the parameters are generated from a prior $\pi(\th)$ then the distribution
of the parameters retained after the rejection process outlined above is given by
\begin{equation}\label{truncated prior}
    \pi_\epsilon(\th)= \frac{\pi(\th) \int_{{\cal B}_\epsilon} f_{\cal M}(\s|\th)d\s}
    {\int_{\Pi} \pi(\th) \int_{{\cal B}_{\epsilon}} f_{\cal M}(\s|\th)d\s d\th}.
\end{equation}

We shall call this density the {\it truncated prior}. It is not hard to check that
\begin{equation}\label{crucial identity}
    \pi(\th|\s_{obs})=\frac{f_{\cal M}(\s_{obs}|\th)\pi(\th)}{\int_{\Pi}f_{\cal M}(\s_{obs}|\th)\pi(\th)d\th}=
    \frac{f_\epsilon(\s_{obs}|\th)\pi_\epsilon(\th)}{\int_{\Pi}f_\epsilon(\s_{obs}|\th)\pi_\epsilon(\th)d\th}.
\end{equation}
Thus the posterior distribution of the model ${\cal M}$ for $\s=\s_{obs}$ given the prior $\pi(\th)$ is exactly
equal to the posterior distribution of the truncated model ${\cal M}_\epsilon(\s_{obs})$ given the truncated prior
$\pi_\epsilon(\th)$. If we can estimate the truncated prior and make an educated guess for a parametric statistical model of
$M_\epsilon(\s_{obs})$, we arrive at a reasonable approximation of the posterior $\pi(\th|\s_{obs})$ even if the
likelihood of the full model ${\cal M}$ is unknown. It is to be expected that due to the localization process the truncated model will exhibit a simpler structure than the full model ${\cal M}$ and thus be easier to estimate.

\medskip

Estimating $\pi_\epsilon(\th)$ is straightforward, at least when the summary statistics can be sampled from ${\cal M}$ in a
reasonable amount of time: Sample the parameters from the prior $\pi(\th)$, create their respective statistics $\s$
from ${\cal M}$ and save those parameters whose statistics lie in ${\cal B}_\epsilon(\s_{obs})$ in a list ${\cal P} = \{
\th^1,\dots,\th^N\}$. The empirical distribution of these retained parameters yields an estimate of $\pi_\epsilon(\th)$.
If the tolerance $\epsilon$ is small then one can assume that $f_{\cal M}(\s|\th)$ is close to some (unknown) constant
over the whole range of ${\cal B}_\epsilon(\s_{obs})$. Under that assumption, formula (\ref{crucial identity}) shows that
$\pi(\th|\s_{obs})\approx\pi_\epsilon(\th)$.
However, when the dimension $n$ of summary statistics is
high -- and for more complex models dimensions like $n=50$ are not unusual --, the ``curse of dimensionality'' raises its ugly head: The tolerance must be chosen rather
large or else the
acceptance rate becomes prohibitively low. This, however, distorts
the precision of the approximation of the posterior distribution by the truncated prior (see \cite{Weg}).
This situation can be partially alleviated by speeding up the
sampling process; such methods are subsumed under
the term {\it approximate Bayesian computation} (ABC).
Marjoram {\it et al.} \cite{Ma}
develop a variant of the classical Metropolis-Hastings algorithm
(termed ABC-MCMC in \cite{Si}) which allows them to sample directly
from the truncated prior $\pi_\epsilon(\th)$. In \cite{Si}
a sequential Monte Carlo sampler (ABC-PRC) is proposed requiring
substantially less iterations than ABC-MCMC. But even when such methods are applied, the assumption that $f_{\cal
M}(s|\th)$ is constant over the $\epsilon$-ball is a very rough one, indeed.

In order to take into account the variation of $f_{\cal
M}(\s|\th)$ within the $\epsilon$-ball, a post-sampling
regression adjustment (ABC-REG) of the sample ${\cal P}$ of retained
parameters is introduced by Beaumont {\it et
al.} \cite{Be}. Basically, they postulate a (locally) linear dependence
between the parameters $\th$ and their associated summary statistics
$\s$. More precisely, the (local) model they implicitly assume is of the form
$\th = {\bf M} \s + {\bf m}_0 + \eps,$
where ${\bf M}$ is a matrix of regression coefficients, ${\bf m}_0$
a constant vector and $\eps$ a random vector of zero mean. Computer
simulations suggest that for many population models ABC-REG yields
posterior marginal densities that have narrower HPD (highest
posterior density) regions and are more closely centered around the
true parameter values than the empirical posterior densities
directly produced by ABC-samplers (\cite{Weg}).
From a point of view of statistical modeling, however, it is unusual to take the parameters ${\th}$ as endogenous and the summary
statistics $\s$ as exogenous variables. As a
consequence, this regression adjustment does not take into account the prior
distribution; it can lead to misspecified posteriors, in particular, it may yield posteriors that are
non-zero in parameter regions where the priors actually vanish (see Figure $1$B for an illustration of this phenomenon).
Moreover, it is not clear how ABC-REG could yield an estimate of the marginal density of model ${\cal M}$ at
$\s_{obs}$, an information that is crucial for model comparison.

\bigskip

One can overcome these drawbacks by stipulating for ${\cal M}_\epsilon(\s_{obs})$ a General Linear
Model (abbreviated as GLM in the literature -- not to be confused with the Generalized Linear Models
which unfortunately share the same abbreviation.) To be precise, we assume the summary statistics $\s$ created by the
truncated model's likelihood $f_\epsilon(\s|\th)$ to satisfy

\begin{equation} \label{GLM}
\s | \th = {\bf C} \th + {\bf c}_0 + \eps,
\end{equation}

where ${\bf C}$ is a $n \times m$-matrix of constants, ${\bf c}_0$ a
$n \times 1$-vector and $\eps$ a random vector with a multivariate
normal distribution of zero mean and covariance matrix $\Sig_s$:
$$ \eps \sim {\cal N}({\bf 0}, \Sig_s). $$

A GLM has the advantage to take into account not only the (local) linearity, but also the
strong correlation normally present between the components of the
summary statistics. Of course, the model assumption (\ref{GLM}) can never represent the full truth since its statistics
are in principle unbounded whereas the likelihood $f_\epsilon(\s|\th)$ is supported on the $\epsilon$-ball
around $\s_{obs}$. But since the multivariate Gaussians will fall off rapidly in practice and not reach far out off the boundary of
${\cal B}_\epsilon(\s_{obs})$ this is to be a disadvantage we can live with. In particular, the OLS-estimate outlined below
implies that for $\epsilon \rightarrow 0$ the constant ${\bf c}_0$ tends to $\s_{obs}$ whereas the design matrix ${\bf C}$
and the covariance matrix $\Sig_s$ both vanish. This means that in the limit of zero tolerance $\epsilon=0$ our model assumption yields
the true posterior distribution of ${\cal M}$.

\subsection*{Theory Section}

In this section we describe the above methodology -- referred to as ABC-GLM in the following -- in more detail. The basic
two-step procedure of ABC-GLM may be summarized as follows:

\begin{description}
\item [GLM1] Given a model ${\cal M}$ creating summary statistics $\s$ and given a value of observed
summary statistics $\s_{obs}$, create a sample of retained parameters
$\th^j,\ j=1,\dots,N$, with aid of some ABC-sampler
(ABC-REJ, ABC-MCMC or ABC-PRC) based on a prior
distribution $\pi(\th)$ and some choice of the tolerance $\epsilon>0$.
\item [GLM2] Estimate the truncated model ${\cal M}_\epsilon(\s_{obs})$ as a General Linear Model
and determine, based on the sample $\th^j$, from the truncated prior $\pi_\epsilon(\th)$
an approximation the posterior $\pi(\th|\s_{obs})$ according to formula (\ref{crucial identity}).
\end{description}

Let us look more closely at these two steps:

\medskip

{\bf GLM1: ABC-sampling.} We refer the reader to \cite{Ma} and
\cite{Si} for details concerning ABC-algorithms and to \cite{MT} for
a comprehensive review of computational methods for genetic data
analysis. Let us just add a few words: Summary statistics tend to be highly
correlated in practice. In that case, the best choice for a distance
function used for the rejection condition $\mbox{dist}(\s,{\bf
s}_{obs})<\delta$ is the Mahalanobis distance (see e.g. \cite{Re}
for a definition). As an alternative, we recommend to reduce the
number of summary statistics by a Principal Components Analysis
(PCA). PCA de-correlates the statistics and the Mahalanobis distance
is then identical to the standard $L^2$-distance. Moreover, PCA has
the advantage of reducing the dimension of the space of summary
statistics. Enough principal components should be retained, though,
to keep the potential loss of information tolerable.
A more sophisticated method of reducing the dimension of summary statistics,
based on Partial Least Squares (PLS), in described in \cite{Weg}.

\medskip

To fix the notation, let ${\cal P} = \{
\th^1,\dots,\th^N\}$ be a sample of vector-valued parameters created
by some ABC-algorithm simulating from some prior $\pi(\th)$,
and ${\cal S}=\{ \s^1,\dots,\s^N\}$ the sample of associated summary
statistics produced by the model ${\cal M}$. Each parameter
${\th^j}$ is an $m$-dimensional column vector
${\th^j}=(\theta^j,\dots,\theta_m^j)^T$ and each summary statistics
an $n$-dimensional column vector $\s^j=(s_1^j,\dots,s_n^j)^T \in {\cal B}_\epsilon(\s_{obs})$. The
samples ${\cal P}$ and ${\cal S}$ can thus be viewed as $m \times
N$- and $n \times N$-matrices ${\bf P}$ and ${\bf S}$, respectively.

The empirical estimate of the truncated prior $\pi_\epsilon(\th)$ is given by the discrete distribution that puts a point mass of $1/N$ on each value $\th^j \in {\cal P}$. We smoothen out this empirical distribution by placing a sharp Gaussian peak over each parameter value $\th^j$. More precisely, we set
\begin{equation}\label{Gauss peaks}
    \pi_\epsilon(\th)=\frac{1}{N}\sum_{j=1}^N \phi(\th-\th^j,\Sig_\theta),
\end{equation}
where
$$
\phi(\th-\th^j,\Sig_\theta)=\frac{1}{|2\pi\Sig_\theta|^{-1/2}} \exp(-\frac{1}{2}(\th-\th^j)^T\Sig_\theta^{-1}(\th-\th^j))
$$
and
$$
\Sig_\theta=\mbox{diag}(\sigma_1,\dots,\sigma_m)
$$
is the covariance matrix of $\phi$ which determines the width of the Gaussian peaks. The larger the number $N$ of sampled parameter values, the sharper the peaks can be chosen in order to still get a rather smooth $\pi_\epsilon$. If the parameter domain $\Pi$ is normalized to $[0,1]^m$, say, then a reasonable choice is $\sigma_k=1/N$. Otherwise, $\sigma_k$ should be adapted to the parameter range of the parameter component $\theta_k$. Too small values of $\sigma_k$ will result in wiggly posterior curves, too large values might  unduly smear out the curves. The best advice is to run the calculations with several choices for $\Sig_\theta$.

\bigskip

{\bf GLM2: General Linear Model.} As explained in the introduction, we assume the truncated model
${\cal M}_\epsilon(\s_{obs})$ to be normal linear, i.e. the random vectors $\s$ satisfy to (\ref{GLM}).
The covariance matrix $\Sig_s$ encapsulates the strong correlations normally
present between the components of the summary statistics. ${\bf C}$,
${\bf c}_0$ and $\Sig_s$ can be estimated by standard multivariate
regression analysis from the sample ${\cal P}$, ${\cal S}$ created
in step {\bf GLM1}\footnote{Strictly speaking, one must redo an ABC-sample from uniform priors over $\Pi$ in order to get an unbiased estimate of the GLM if the prior $\pi(\th)$ is not uniform already. On the other hand, ordinary least squares estimators are quite insensitive to the prior's influence. In practice, one can as well use the sample ${\cal P}$ to to the estimate.} To be specific, set ${\bf X}=({\bf 1} | {\bf
P}^t)$, where ${\bf 1}$ is an $N \times 1$-vector of 1's. ${\bf C}$
and ${\bf c}_0$ are determined by the usual least squares estimator

$$
(\hat{{\bf c}}_0 | \hat{{\bf C}})={\bf S} {\bf X} ({\bf X}^t {\bf
X})^{-1},
$$

and for $\Sig_s$ we have the estimate

$$
\hat{\Sig}_s=\frac{1}{N-m} \hat{{\bf R}}^t \hat{{\bf R}},
$$

where $\hat{{\bf R}}={\bf S}^t-{\bf X} \cdot (\hat{{\bf c}}_0 |
\hat{{\bf C}})^t$ are the residuals. The likelihood for this model -- dropping the hats on the matrices to simplify the notation -- is
given by

\begin{equation}\label{likelihood}
f_\epsilon(\s|\th)=| 2\pi \Sig_s|^{-1/2} \cdot \exp\left(-\frac{1}{2} (\s -
{\bf C} \th -{\bf c}_0)^t \ \Sig_s^{-1} \ (\s - {\bf C} \th -{\bf
c}_0) \right).
\end{equation}

An exhaustive treatment of linear models in a Bayesian (econometric)
context is given in Zellner's book \cite{Ze}.

\bigskip

Recall from (\ref{crucial identity}) that for a prior
$\pi(\th)$  and an observed summary statistic $\s_{obs}$, the
parameter's posterior distribution for our full model ${\cal M}$ is given by

\begin{equation} \label{post1}
\pi(\th |\s_{obs}) = c \cdot f_\epsilon(\s_{obs}|\th) \pi_\epsilon(\th).
\end{equation}

where $f_\epsilon(\s_{obs}|\th)$ is the likelihood of the truncated model ${\cal M}_\epsilon(\s_{obs})$ given by (\ref{likelihood}) and $\pi_\epsilon(\th)$ is the estimated (and smoothed) truncated prior given by (\ref{Gauss peaks}).

Performing some matrix acrobatics (see e.g. the proof of Lemma 2.1 in \cite{Li}) one can show that the posterior (\ref{post1}) is -- up to a multiplicative constant -- of the form $\sum_{i=j}^N \exp(-\frac{1}{2}Q_j)$ where

\begin{eqnarray*}\label{Q_j}
    Q_j &=&(\th-{\bf t}^j)^T {\bf T}^{-1}(\th-{\bf t}^j)+(\s_{obs}-{\bf c}_0)^T\Sig_s^{-1}(\s_{obs}-{\bf c}_0)+\dots\\
    && \ldots + (\th^j)^T\Sig_\theta^{-1}\th^j - ({\bf v}^j)^T{\bf T}{\bf v}^j.
\end{eqnarray*}

Here ${\bf T}, {\bf t}^j$ and ${\bf v}^j$ are given by

\begin{equation}\label{T}
{\bf T} = \left( {\bf C}^t \Sig_s^{-1} {\bf C} + \Sig_\theta^{-1}
\right)^{-1}
\end{equation}

and ${\bf t}^j= {\bf T} {\bf v}^j$, where

\begin{equation}\label{v}
    {\bf v}^j =  {\bf C}^t \Sig_s^{-1} (\s_{obs} - {\bf c}_0)+ \Sig_\theta^{-1}
    \th^j.
\end{equation}

From this we get

\begin{equation}\label{post2}
    \pi(\theta|\s_{obs}) \propto
    \sum\limits_{j=1}^N c(\th^j) \exp\left( - \frac{1}{2}(\th -
    {\bf t}^j)^T {\bf T}^{-1} (\th -
    {\bf t}^j) \right),
\end{equation}
where
\begin{equation}\label{c}
    c(\th^i)=\exp \left[ -\frac{1}{2} \left( (\th^j)^t \Sig_\theta^{-1}
    \th^j - ({\bf v}^j)^T {\bf T} {\bf v}^j \right)
    \right].
\end{equation}

When the number of parameters exceeds two, graphical visualization
of the posterior distribution becomes impractical and marginal
distributions must be calculated. The marginal posterior density of
the parameter $\theta_k$ is defined by

$$
\pi(\theta_k|\s)=\int_{\R^{m-1}} \pi(\th | \s) d\th_{-k},
$$
where integration is performed along all parameters except
$\theta_k$.

Recall that the marginal distribution of a multivariate normal ${\cal N}(\mmu,\Sig)$ with respect to the $k$-th component is the univariate normal density ${\cal N}(\mu_k,\sigma_{k,k})$.
Using this fact, it is not hard to show that the marginal posterior of parameter $\th_k$ is given by
\begin{equation}\label{marginal posterior}
\pi(\theta_k|\s_{obs}) = a \cdot
    \sum\limits_{j=1}^N c(\th^j) \exp\left( - \frac{(\theta_k -
    t_k^j)^2}{2\tau_{k,k}} \right).
\end{equation}
where $\tau_{k,k}$ is the $k$-th diagonal element of the matrix ${\bf T}$, $t_k^j$ is the $k$-th component of the vector ${\bf t}^j$, and $c(\th^j)$ is still determined according to (\ref{c}). The normalizing constant $a$ could, in principle, be determined analytically but is in practice more easily recovered by a numerical integration. Strictly speaking, the integration should only be done over the bounded parameter domain $\Pi$ and not over the whole of $\R^m$. But this no longer allows for an analytic form of the marginal posterior distribution. For large values of $N$ the diagonal elements in the matrix $\Sigma_\theta$ can be chosen so small that the error is in any case negligible.

\bigskip

{\bf Model selection.} The principal difficulty of model selection methods in non-parametric
settings is that it is nearly impossible to estimate the likelihood of ${\cal M}$ at $\s_{obs}$ due to the
high dimension of the summary statistics (``curse of dimensionality''); see \cite{Be2} for an
approach based on multinomial logit. Parametric models on the other hand lend themselves readily
to model selection via Bayes factors. Given the model ${\cal M}$, one must determine the marginal density
$$
f_{\cal M}(\s_{obs})=\int_{\Pi} f(\s_{obs}|\th) \pi(\th) d\th.
$$
It is easy to check from (\ref{truncated likelihood}) and (\ref{truncated prior}) that
$$
f_{\cal M}(\s_{obs})=A_\epsilon(\s_{obs},\pi) \cdot \int_{\Pi} f_\epsilon(\s_{obs}|\th) \pi_\epsilon(\th) d\th.
$$

Here
\begin{equation}\label{acceptance rate}
A_\epsilon(\s_{obs},\pi):=\int_{\Pi} \pi(\th) \int_{{\cal B}_{\epsilon}} f_{\cal M}(\s|\th)d\s d\th
\end{equation}

is the acceptance rate of the rejection process. It can easily be estimated with aid of ABC-REJ: Sample parameters from the prior $\pi(\th)$, create the corresponding statistics $\s$ from ${\cal M}$ and count what fraction of the statistics fall into the $\epsilon$-ball ${\cal B}_\epsilon$ centered at $\s_{obs}$.

If we assume the underlying model of ${\cal M}_\epsilon(\s_{obs})$ to be our GLM then the marginal
density of ${\cal M}$ at $\s_{obs}$ can be estimated as follows:
\begin{equation}\label{marginal density type C}
f_{\cal M}(\s_{obs})=\frac{A_\epsilon(\s_{obs},\pi)}{N |2\pi {\bf D}|^{1/2}} \sum\limits_{j=1}^N \exp\left(-\frac{1}{2} (\s_{obs}-{\bf m}^j)^T {\bf D}^{-1} (\s_{obs}-{\bf m}^j)\right)
\end{equation}
where the sum runs over the parameter sample ${\cal P}= \{
\th^1,\dots,\th^N\}$,
$$
{\bf D}=\Sig_s+{\bf C}\Sig_\theta{\bf C}^T
$$
and
$$
{\bf m}^j={\bf c}_0 + {\bf C} \th^j.
$$

For two models ${\cal M}_A$ and ${\cal M}_B$ with prior probabilities $\pi_A$ and $\pi_B=1-\pi_A$,
the Bayes factor $B_{AB}$ in favor of model ${\cal M}_A$ over model ${\cal M}_B$ is
\begin{equation}\label{bayes factor}
B_{AB}=\frac{f_{{\cal M}_A}(\s_{obs})}{f_{{\cal M}_B}(\s_{obs})}
\end{equation}
where the marginal densities $f_{{\cal M}_A}$ and $f_{{\cal M}_B}$ are calculated
according to (\ref{marginal density type C}). The posterior probability of model ${\cal M}_A$ is
$$
f({\cal M}_A|\s_{obs}) = \frac{B_{AB}\pi_A}{B_{AB}\pi_A + \pi_B}.
$$

\bigskip

\subsection*{Simulation Section}

\begin{figure}[h!]
\centering
\includegraphics[width=\textwidth]{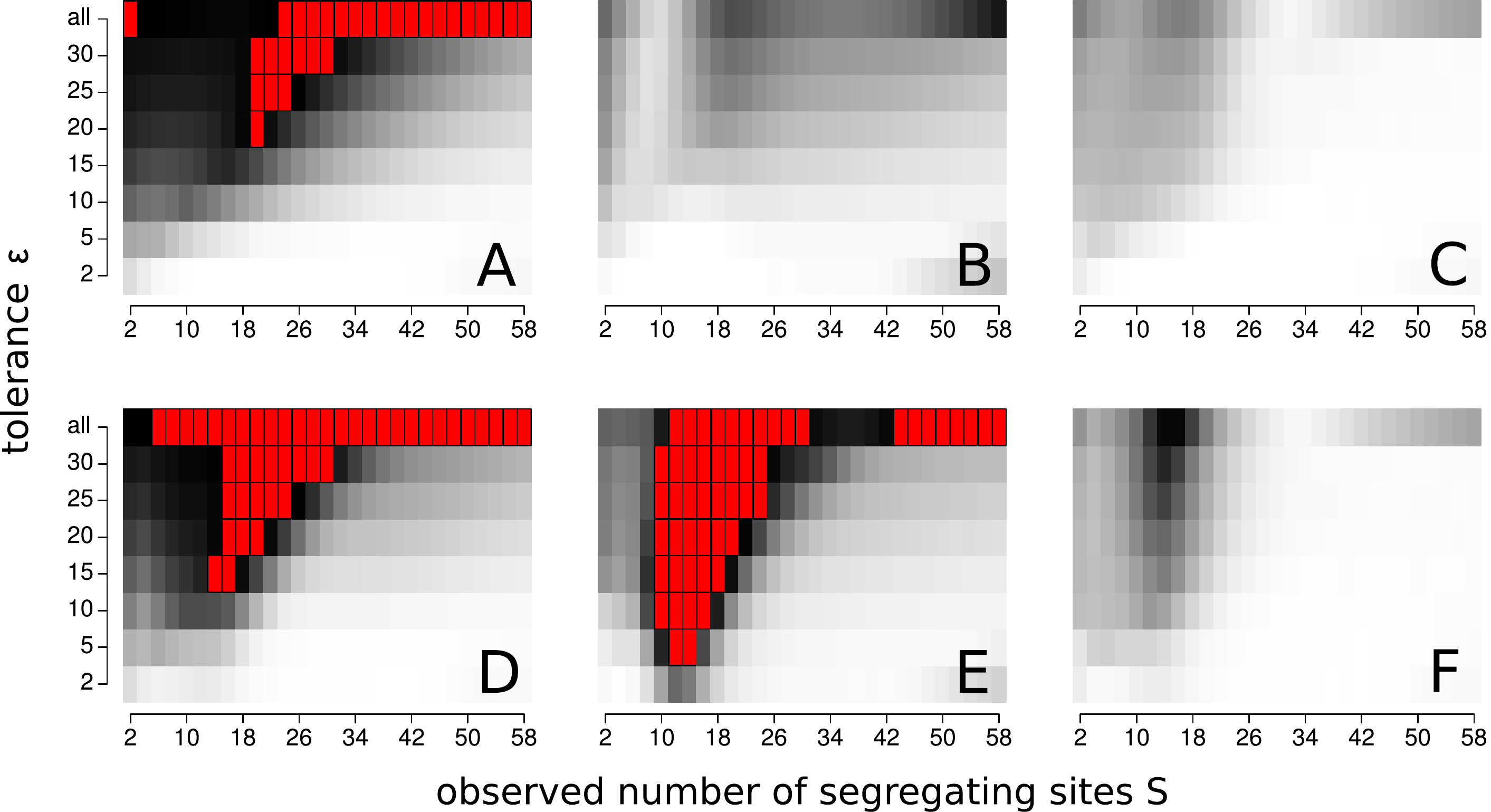}
\caption{\small Comparison of rejection (A and D), ABC-REG (B and E) and ABC-GLM (C and F) posteriors with those obtained from analytical likelihood calculations. Shades indicate the distance between inferred and analytically calculated posterior (see text). White corresponds to an exact match (zero distance) and darker grey shades indicate larger distances. If the the inferred posterior differs from the analytical more than the prior does, squares are marked in red. Shown are averages over 25 independent estimations.}
\label{tavare_comparison}
\end{figure}

{\bf A toy model.}
In Figure \ref{tavare_comparison} we present the comparison of posteriors obtained with rejection sampling, ABC-REG and ABC-GLM, with those determined analytically (``true posteriors''). As a toy model we inferred the population-mutation parameter $\theta=4N\mu$ from the number of segregating sites $S$ of a sample of sequences with 10'000 bp for different observed values and tolerance levels. Estimations are always based on 5000 simulations with $\mbox{dist}(S,S_{obs})<\epsilon$, and we report the average of 25 independent replications per data point. Estimation bias of the different approaches was assessed by computing the $L_1$-distance between the inferred posterior and the true one obtained from analytical calculations using the likelihood function introduced by Watterson \cite{Wa}. Recall that the $L_1$-distance of two (continuous) densities $f(\theta)$ and $g(\theta)$
$$
\|f-g\|_1 = \int |f(\theta)-g(\theta)|d\theta
$$
measures the area between the function curves. It is equal to $2$ when $f$ and $g$ have disjoint supports and it vanishes when the functions are identical.

When we used a uniform prior $\theta \sim \mbox{Unif}([0.005, 10])$ (Figure \ref{tavare_comparison}A to C), both ABC-REG and ABC-GLM give comparable results and improve the posterior estimation compared to the simple rejection algorithm except for very low tolerance values $\epsilon$ where the rejection algorithm is expected to be very close to the true posterior. The average $L_1$-distance over all observed data sets and tolerance values $\epsilon$ are 0.4726, 0.2609 and 0.1814 for the rejection algorithm, ABC-REG and ABC-GLM, respectively. Note that perfect matches between the approximate and the true posteriors are difficult to obtain because all approximated posteriors depend on a smoothing step which may not give accurate results close to boarders of their supports. In order to have a fair comparison, we adjusted the smoothing parameters (bandwidths) such as to get the best results for both approaches.

\begin{figure}[h!]
\centering
\includegraphics[width=\textwidth]{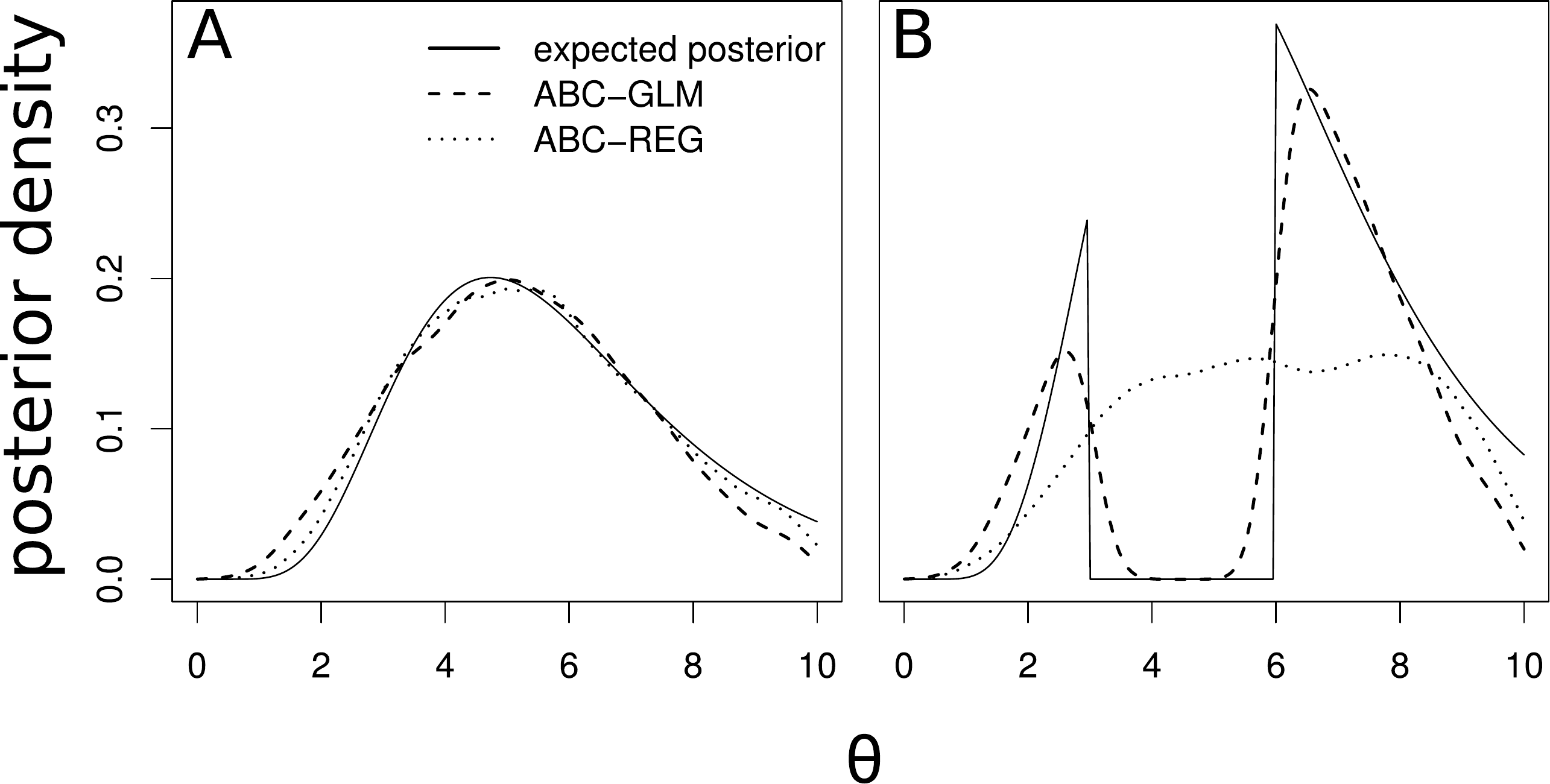}
\caption{\small Posterior estimates using ABC-GLM and ABC-REG for $S_{obs}=16$ based on 5000 simulations with $\mbox{dist}(S,S_{obs})<10$ using a discontinuous prior $\theta \sim \mbox{Unif}([0.005, 3] \cup [6, 10])$. }
\label{example}
\end{figure}

However, when we used a discontinuous prior $\th  \sim \mbox{Unif}([0.005, 3] \cup [6, 10])$ with an -- admittedly quite artificial -- ``gap'' in the middle, we observed a quite distinct pattern (see Figure \ref{example} and Figure \ref{tavare_comparison}D to E). One clearly recognizes that posteriors inferred with ABC-REG are frequently misplaced and often even further away from the true posterior (in $L_1$-distance) than the prior. The reason for this is that in the regression step of ABC-REG parameter values may easily be shifted outside the prior support. This behavior of ABC-REG has been observed earlier and as an \textit{ad hoc} solution Hamilton \textit{et al.} \cite{Ha} proposed to transform the parameter values prior to the regression step by a transformation of the form $y=-ln(tan(\dfrac{x-a}{b-a}\dfrac{\pi}{2})^{-1})$ where $a$ and $b$ are the lower and upper borders of the prior support interval. For more complex priors -- like the discontinuous prior used here -- this transformation may not work. In fact, we applied this transformation whenever we performed ABC-REG. By adding additional borders, smoothing introduces a larger bias (see Figure \ref{example}). Still, ABC-GLM is much less affected by the ``gap'' prior than ABC-REG. The average $L_1$-distance over all observed data sets and tolerance values $\epsilon$ are 0.4412, 0.4925 and 0.1876 for the rejection algorithm, ABC-REG and ABC-GLM, respectively.

Example posteriors with $S_{obs}=16$ based on 5000 simulations with $\mbox{dist}(S,S_{obs})<10$ are shown in Figure \ref{example}. While the two approximated posteriors (ABC-REG and ABC-GLM) are very similar when the priors are uniform over the whole range (Figure \ref{example}A), they become markedly different when we use a uniform prior with a ``gap'' between $3$ and $6$ (Figure \ref{example}B). Note that the posterior estimated with ABC-REG is maximal in the forbidden region! This phenomenon clearly illustrates that ABC-REG is not consistent with the prior distribution.

{\bf Application to chimpanzees.}
In standard taxonomies, chimpanzees, the closest living relatives of humans, are classified into two species: the common chimpanzee (\textit{Pan troglodytes}) and the bonobo (\textit{Pan paniscus}). Both species are restricted to Africa and diverged roughly $9$ MYA ago \cite{Won, Bec1}. The common chimpanzees are further subdivided into three large populations or subspecies based on their separation by geographic barriers. Among them, the western chimpanzees (\textit{P. troglodytes verus}) form the most remote group. Interestingly, recent multilocus studies found consistent levels of gene flow between the western and the central (\textit{P. t. troglodytes}) chimpanzees \cite{Won, Bec1}. Nonetheless, a recent study of 310 microsatellites in 84 common chimpanzees supports a clear distinction between the previously labeled populations \cite{Bec2}. Using a PCA analysis, indication for substructure within the western chimpanzees was found in the same study.

To demonstrate the applicability of the model selection given in the Theory Section we contrast two different models of the western chimpanzee population with this data set: a model of a single panmictic population with constant size and a finite island model of constant size and constant migration among demes. While we estimated $\theta=2 N_e \mu$, priors were set on $N_e$ and $\mu$ separately with $log_{10}(N_e) \sim \mbox{Unif}([3,5])$ and $\mu \sim N(5\cdot 10^{-4}, 2 \cdot 10^{-4})$ truncated on $\mu \in [10^{-4},10^{-3}]$. In the case of the finite island model we had an additional prior $n_{pop} \sim \mbox{Unif}([10, 100])$ on the number of islands, and individuals were attributed randomly to the different islands.

We obtained genotypes for all $50$ individuals reported to be of western chimpanzee origin  from the study of Becquet \textit{et al.} \cite{Bec2} excluding captive born hybrids. We checked the mutation pattern for each individual locus, and all alleles not matching the pattern were set as missing data. A total of $265$ loci \cite{Bec2} were used, after removing the loci on the X and Y chromosome as well as those being monomorphic among the western chimpanzees. All simulations were performed using the software SIMCOAL2 \cite{La} and we reproduced the pattern of missing data observed in the dataset. Using the software package {\it Arlequin3.0} \cite{Ex} we calculated two summary statistics on the dataset: the average number of alleles per locus, $K$, and $F_{IS}$, the fixation index within the western chimpanzees. We performed a total of 100,000 simulations per model.

\begin{figure}[h!]
\includegraphics[width=6.5cm]{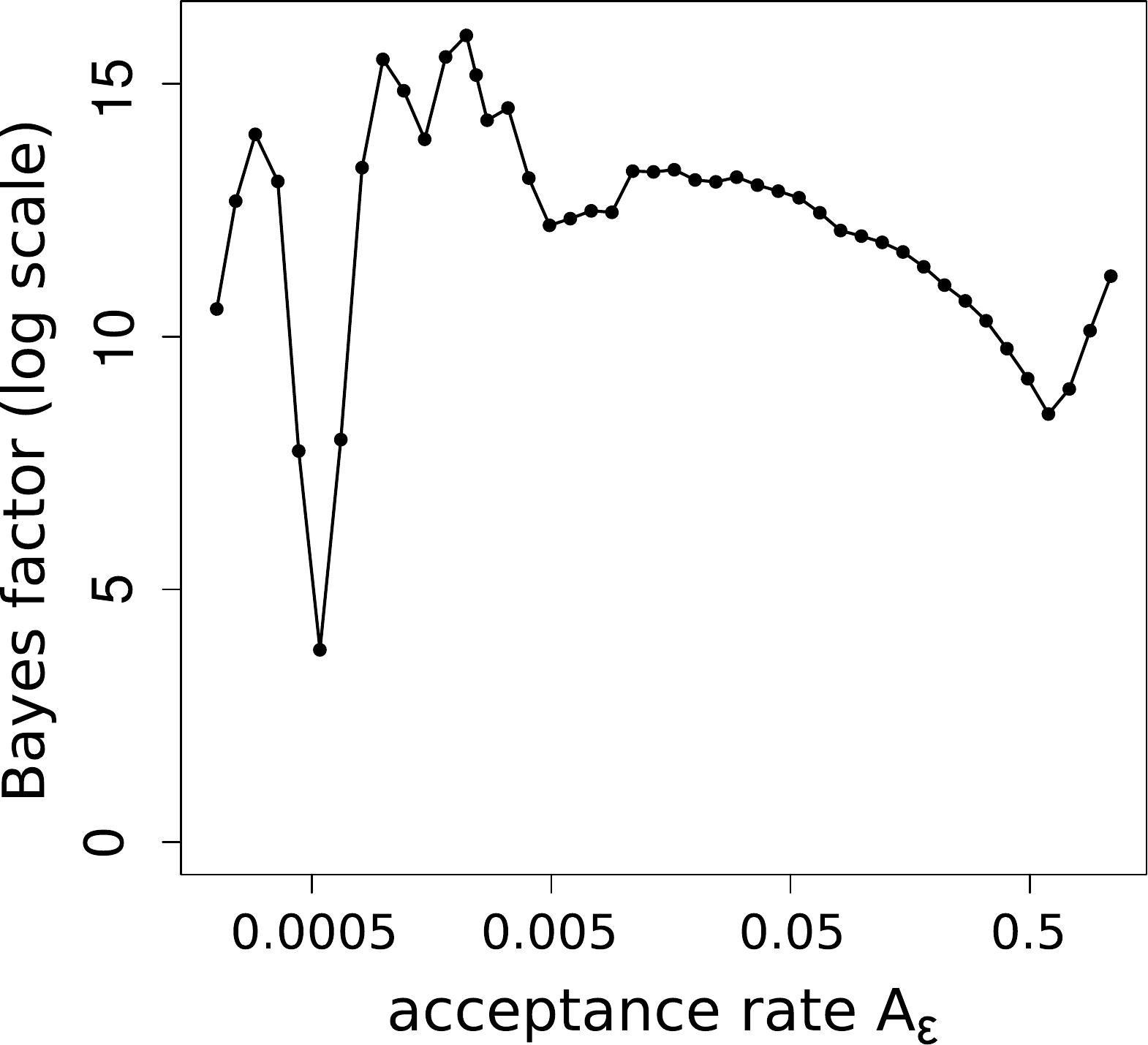}
\caption{\small Bayes factor for the island relative to the panmictic population model for different acceptance rates (logarithmic scale). For very low acceptance rates we observe large fluctuations whereas the Bayes factor is quite stable for larger values. Note that $A_\epsilon \leq 0.005$ corresponds to $\leq 500$ simulations, too small a sample size for trustful statistical model estimation.}
\label{Bayes factor}
\end{figure}

In Figure \ref{Bayes factor} we report the Bayes factor of the island model according to (\ref{Bayes factor}) for different acceptance rates $A_\epsilon$, see (\ref{acceptance rate}). While there is a large variation for very small acceptance rates, the Bayes factor stabilizes for $A_\epsilon \geq 0.005$. Note that $A_\epsilon \leq 0.005$ corresponds to less than $500$ simulations, and that the ABC-GLM approach, based on a model estimation and a smoothing step, is expected to produce poor results since the estimation of the model parameters is unreliable due to the small sample size. The good news is that the Bayes factor is stable over a large range of tolerance values. We may therefore safely reject the panmictic population model in favor of population subdivision among western chimpanzees with a Bayes factor of $B \approx e^{12}>10^{5}$.

\smallskip

\subsection*{Discussion}
Due to still increasing computation power it is nowadays possible to tackle estimation problems in a Bayesian framework for which analytical calculation of the likelihood is inhibited. In such cases, Approximate Bayesian Computations are often the choice. A key innovation in speeding up such algorithms was the use of a regression adjustment, termed ABC-REG in this note, which used the frequently present linear relationship between generated summary statistics ${\bf s}$ and parameters of the model $\th$ in a neighborhood of the observed summary statistics ${\bf s_{obs}}$ \cite{Be}. The main advantage is that larger tolerance values $\epsilon$ still allow to extract reasonable information about the posterior distribution $\pi(\th |{\bf s})$ and hence less simulations are required to estimate the posterior density.

Here we present a new approach to estimate approximate posterior distributions, termed ABC-GLM, similar in spirit to ABC-REG, but with two major advantages: First, by using a GLM to estimate the likelihood function, ABC-GLM always consistent with the prior distribution. Secondly, our ABC-GLM approach is naturally embedded into a standard Bayesian framework, which in turn allows the application of well-known Bayesian methodologies such as model averaging or model selection via Bayes factors.

ABC-GLM is compatible with any type of ABC-sampler, including likelihood-free MCMC \cite{Ma}, \cite{Be3}. Also, more complicated regression regimes taking non-linearity or heteroscedacity into account may be envisioned when the GLM is replaced by some more complex model. A great advantage of the current GLM-setting is its simplicity which renders implementation in standard statistical packages feasible.

{\bf Application to chimpanzees.}
We showed the applicability of the model selection procedure via Bayes factors by opposing two different models of population structure among the western chimpanzees \textit{Pan troglodytes verus}. Our analysis strongly suggests population substructure within the western chimpanzees since an island model is significantly favored over a model of a panmictic population. While none of our simple models is thought to mimic the real setting exactly (models never do!), we still believe that they capture the main characteristics of the demographic history influencing our summary statistics, namely the number of alleles $K$ and the fixation index $F_{IS}$. While the observed $F_{IS}$ of 2.6\% has been attributed to inbreeding previously \cite{Bec2}, we propose that such values may easily arise if diploid individuals are sampled in a randomly scattered way over a large, substructured population. While it was almost impossible to simulate the value $F_{IS}=2.6\%$ in the model of a panmictic population, it easily falls within the range of values obtained from an island model.

\bigskip

{\bf Acknowledgements.} The authors would like to thank David J. Balding and Christian P. Robert for their useful comments on a first draft of this preprint.

\end{document}